# A global unstructured, coupled, high-resolution hindcast of waves and storm surges

Lorenzo Mentaschi, Michalis Vousdoukas, Guillermo Garcia-Sanchez, Tomas Fernandez Montblanc, Aron Roland, Evangelos Voukouvalas, Ivan Federico, Ali Abdolali, Yinglong J. Zhang, Luc Feyen

## Abstract

Accurate information on waves and storm surges is essential to understand coastal hazards that are expected to increase in view of global warming and rising sea levels. Despite the recent advancement in development and application of large-scale coastal models, nearshore processes are still not sufficiently resolved due to coarse resolutions, transferring errors to coastal risk assessments and other large-scale applications. Here we developed a 50-year hindcast of waves and storm surges on an unstructured mesh of >650,000 nodes with an unprecedented resolution of 2-4 km at the global coast. Our modelling system is based on the circulation model SCHISM that is fully coupled with the WWM-V (WindWaveModel) and is forced by surface winds, pressure, and ice coverage from the ERA5 reanalysis. Results are compared with observations from satellite altimeters, tidal gauges and buoys, and show good skill for both Sea Surface Height (SSH) and Significant Wave Height ($H_s$), and a much-improved ability to reproduce the nearshore dynamics compared with previous, lower-resolution studies. Besides SSH, the modelling system also produces a range of other wave-related fields at each node of the mesh with a time step of 3 hours, including the spectral parameters of the first three largest energy peaks. This dataset offers the potential for more accurate global-scale applications on coastal hazard and risk.

## Introduction

Waves and storm surges are complex oceanographic phenomena that can have significant impacts on coastal communities, particularly during extreme weather events such as hurricanes, typhoons, and extratropical cyclones. Global warming is expected to increase extreme sea levels and coastal flooding around the world (McGranahan et al., 2007; Wahl et al., 2017; Vousdoukas et al., 2018; Tebaldi et al., 2021). To better understand trends in past and future coastal hazards and risks there is a growing need for accurate modeling of waves and storm surge at a global scale, incorporating the mutual impact of wind-waves and storm surge models in a coupled framework. In the last decades, advances in numerical modeling, computing power and availability of coastal observations (in-situ and remote sensing) have enabled the development of global models of waves and storm surges that can provide valuable information for coastal hazard assessment and risk management (e.g. Vousdoukas et al., 2020). The skill of existing large scale coastal hindcasts and reanalyses is generally good when compared versus satellite observations offshore. Closer to shores, however, complex coastal and bathymetric features and interactions hamper the ability of existing models to fully capture the relevant dynamics, partly due to their low resolution and the lack of coupling waves and circulation. In addition, the dynamic interaction of coastal circulation, wind wave, inland hydrology, ice and atmospheric models are not fully captured in standalone or 1 way coupled fashion (Moghimi et al., 2020; Abdolali et al., 2021).

In recent years, unstructured-grid models are more and more becoming an alternative to regular grids for large-scale simulations. These models are flexible by employing meshes with varying resolution that can be adjusted to fit any arbitrary geometry, making them ideal for modeling coastal environments with irregular coastlines, complex bathymetry and structures. The history of unstructured wave and storm surge models dates back to the 1990s when researchers began exploring the use of finite element and finite volume numerical methods to solve differential equations (e.g. Luettich et al., 1992; Lynch et al., 1996), and they soon became a consolidated tool for coastal modelling. Nowadays, among the established circulation unstructured models are ADCIRC (Luettich et al., 1992; Pringle et al., 2021), the Finite-Volume Coastal Ocean Model (FVCOM, Chen et al., 2003), the Semi-implicit Cross-scale Hydroscience Integrated System Model (SCHISM, Y. Zhang & Baptista, 2008; Y. J. Zhang et al., 2016), the System of HydrodYnamic Finite Element Modules (SHYFEM, Umgiesser and Zampato, 2001; Micaletto et al., 2022), the model TELEMAC (Hervouet and Bates, 2000).

Among the earliest spectral wave models employing numerical schemes unstructured grids were the model TOMAWAC (Benoit et al., 1997) and CREST e.g. Ardhuin & Herbers, 2005), which solve the wave action balance equation (WAE) based on semi-Lagrangian schemes. In the following other models developed, which solve the WAE based on a Eulerian viewpoint, in analogy to the mostly applied models based on structured grids (WAM, WW3, SWAN), which have been WWM (Liau, 2001, Hsu et al., 2005, Roland et al. 2006) and MIKE2-SW (Soerensen et al., 2004).

In the following decade the development of the Eulerian-approach gained further momentum and mostly all of the available spectral wave models have been enhanced for the possibility to apply unstructured grids (e.g. Roland, 2008 (WWM-II), 2012b (WAM), Zijlema et al. 2008 (UnSWAN), Qi et al.2009 (FVCOM).

The schemes developed in Roland, 2008 included efficient implicit schemes to overcome the strict CFL constraint and resulted in WWM-II, which was fully coupled to SCHISM and parallelized using Domain Decomposition methods and lead to WWM-III, Roland et al., 2012a). The latest efforts in Sikiric et al. 2018 and further development in this study resulted in the latest Version of the wind wave model (WWM-V), which was utilized in this study. We have validated the implicit schemes in WWM-V and compared those to the latest WW3 version, which showed only marginally worse performance for the global setup. Furthermore, the integration of the triads by applying Patankar rules and underrelaxation of the wave breaking and triad source term to increase the robustness of the model have improved in the implicit scheme. Moreover, the memory management in the model was improved in terms of memory alignment, memory usage and cache locality, which improved the performance of the model in contrast to WWM-III.

Most of the elements have been implemented in the actual "development" Version of the WW3 Framework. The implementation of implicit solvers as done helped provided an opportunity to further develop the wave models to resolve the dominant coastal processes (Abdolali et al., 2020).

For long, the use of unstructured meshes was confined to local high-resolution studies that deepened insights into the complexity of nearshore interactions (e.g. Roland et al., 2008; Roland & Ardhuin, 2012, Bertin et al., 2012; Federico et al., 2017; Amores et al., 2020; Park et al., 2022). Large/global scale unstructured models of circulation and storm surges were developed in the late 2010s (e.g. Muis et al., 2016; Vousdoukas et al., 2018; Fernández-Montblanc et al., 2020; Saillour et al., 2021). Recently, Zhang et al., (2023) developed a global 3D baroclinic model that includes tides using SCHISM. Existing global studies on circulation and storm surges can reach high spatial resolution at large scale. For example, Wang et al., (2022) reaches a global resolution of about 1.25 km. Zhang et al., (2023) have an offshore resolution of 10-15 km, and nearshore of 3 km globally and 1-2 km in North America.

For waves, the adoption of unstructured meshes at large scale was slower, related to the high computational demand of spectral wave models. The first study using unstructured local subgrids in a global wave model was (Rascle and Ardhuin, 2013). More recent applications pioneered the viability of global unstructured wave models (Mentaschi et al., 2020; Brus et al., 2021), though retaining a relatively low resolution at the global coast of ~10 km. This was achieved thanks to the adoption of efficient domain decomposition techniques and governance of the CFL number (Abdolali et al., 2020), and of subscale modelling techniques that simplify the workload of mesh generation by introducing a parameterization of wave dampening due to unresolved features (Mentaschi et al., 2015b, 2018). Also worth mentioning, there are studies on large basins like the Mediterranean Sea, reaching very high spatial resolution in the order of 100 m, carried out with WW3 and WWM (Lira-Loarca et al., 2022; Toomey et al., 2022).

Another important aspect of ocean modelling is the coupling between waves and circulation. Wind waves can alter circulation in the ocean and affect the surface roughness and near-surface turbulence, while currents can influence the propagation and characteristics of waves, via Doppler effect, whitecapping when propagating against counter currents and by fostering non-linear interactions. Accounting for this coupling is important for accurate modelling, especially in nearshore environments. The coupling of waves and circulation was explored in a number of local studies (e.g. Ferrarin et al., 2008; Roland et al., 2009, 2012; Babanin et al., 2011; Staneva et al., 2016; Clementi et al., 2017; Causio et al., 2021), and only recently global applications were developed (e.g. Law-Chune et al., 2021). Incorporation of variability of water level (tide + surge) in the wave model and inland hydrology enhance the model accuracy in nearshore and provided better estimates of wetting and drying and compound flooding (Bakhtyar et al., 2020; Moghimi et al., 2020).

Taking advantage of these recent advances, here we developed a global coupled model that simulates waves and storm surges on an unstructured mesh with an unprecedented resolution at the coasts of 2-4 km globally. The coupled model combines the circulation model SCHISM (Zhang et al., 2016) and the spectral wave model WWM (Roland, 2008) and is forced by the latest-generation atmosphere reanalysis ERA5 (Hersbach et al., 2020). With this setup, we carried out a 53-year hindcast of waves and storm surges and present here its validation versus satellite and field observations, and a comparison with results from prior lower-resolution studies.

## Materials and Methods

### Model setup

Our modelling framework consists of the Semi-implicit Cross-scale Hydroscience Integrated System Model (SCHISM, Zhang et al., 2016, 2023) that is 2-way coupled with the 3[rd]-generation spectral wave model Wind Wave Model (WWM-V, Roland et al., 2012; Dutour Sikirić et al., 2018).

SCHISM is an extension of the original SELFE code (Y. Zhang & Baptista, 2008) that has undergone several enhancements and upgrades, including the inclusion of large-scale eddying and a seamless cross-scale capability from small bodies of water to the ocean. The model solves the Navier-Stokes equation, under hydrostatic and Boussinesq approximation, on unstructured grids, using a Finite-Element semi-implicit Galerkin scheme for horizontal motions, and a Finite-Volume scheme for vertical motions. The model accounts for the combined effects of wind and atmospheric pressure gradients. In this study, SCHISM was configured in its two-dimensional barotropic mode, which increases model efficiency and is a reasonable approximation for global applications with high computational and storage requirements (Fernández-Montblanc et al., 2019). In this work, SCHISM was used to simulate the generation and propagation of storm surges. The surface stress in SCHISM was calculated using a bulk formula with the drag coefficient computed estimated as a function of the wave motion. The bed shear stress was computed using the Manning approach, assuming a value of 0.02 for the Manning friction coefficient. A biharmonic viscosity filter with a diffusion number-like dimensionless constant of 0.025 is used to suppress grid-scale noises.

WWM-V is a 3[rd]-generation state-of-the-art spectral wave model that solves the wave action equation on unstructured grids using the CRD-N-Scheme (e.g Ricchiuto et al. 2005) for the advection part and discreitzing the spectral part using simple $1^{st}$ order implicit finite difference schemes, rending the whole approach monotone and convergent. The source terms have been linearized (e.g. Patankar, 1980) and if the linearization was not possible, we have underrelaxed the source terms as done for the triad-wave-wave interactions source term. For this study we employed the fully implicit RD-schemes with no splitting, using a Block-Gauss & Jacobi iterative solvers (e.g.: Abdollali, 2020) on unstructured grids. The spectral domain used in this study consists of 36 frequencies and 24 directions. The frequencies are separated by a constant factor, from 0.04 Hz, corresponding to a period of 25 s, to 1 Hz, corresponding to a period of 1 second. The growth/dissipation of the waves was parameterized using (Ardhuin et al., 2010), setting the parameter BETAMAX to the value suggested for ECMWF data, while the non-linear interactions are reproduced by means of the Discrete Interaction Approximation (DIA) (Hasselmann and Hasselmann, 1985). With a resolution of 2 km along coasts, shallow-water interactions can play a significant role. In particular, the bottom friction was parameterized using the JONSWAP approximation (Hasselmann et al., 1973). Depth-limited breaking was represented using the approach by (Battjes and Janssen, 1978).In particular, the Unresolved Obstacles Source Term (UOST, Mentaschi, Pérez, et al., 2015; Mentaschi et al., 2018, 2019, 2020) was introduced to parameterize the energy dampening of small unresolved islands. Furthermore, as the ice plays an important role on wave dynamics at high latitudes, an approach based on UOST was implemented to dampen the energy as a function of the ice concentration. No treatment of ice was adopted in the circulation model. Finally, the methodology proposed by (Tracy et al., 2006) for spectral decomposition was implemented in the model. This approach allows for saving the mean parameter for the separate peaks of the wave spectrum, providing summarized information on the spectra at each node.

The models are 2-way coupled, in that the hydrodynamic component provides the wave model with current velocities and sea levels, while the sea surface roughness is estimated by WWM-V and passed back to SCHISM.

The hindcast simulation was carried out on an unstructured mesh of 651,652 nodes covering the global ocean with a resolution ranging from about 50 km offshore to about 2-4 km nearshore globally. The mesh was generated using the OceanMesh2D tool (Roberts et al., 2019). The bathymetry was taken from the General Bathymetric Chart of the Oceans (GEBCO) dataset (Weatherall et al., 2015). The model was forced with hourly data of sea level pressure, wind speed and sea ice concentration from the fifth-generation reanalysis ERA5 (Hersbach et al., 2020). Apart from its resolution of 25 km, high for a global reanalysis, we selected ERA5 for its extended time span (from 1950 on), and its accuracy, compared to other products, in reproducing long-term trends (Erikson et al., 2022). The model time step was 400 s for both the circulation component and the wave component, with circulation-wave coupling occurring at each time step. The time span of the simulation is from 1970 to 2022. Model output was saved at each domain node with a time frequency of 3 hours for the variables listed in Table 1.

*Table 1: List of variables saved at each mesh node with a 3-hour time step.*

| Name | Description | Units |
|------|-------------|-------|
| Sea Surface Height | Residual water elevation (tides are not included) | m |
| Wind Speed | Wind speed at 10 m (from ERA5) | m s$^{-1}$ |
| Currents Velocity | Horizontal water velocity | m s$^{-1}$ |
| Significant Wave Height | Significant Wave Height | m |
| Mean Period $T_{0-1}$ | Mean period weighted on the spectral component energy | s |
| Zero-Crossing Period $T_{0-2}$ | Zero-crossing mean period | s |
| Overtopping/runup period $T_{10}$ | Mean period for overtopping/runup | s |
| Mean Wavelength | Mean Wavelength | m |
| Mean Wave Direction | Mean Wave Direction, oceanographic convention | °N |

| Mean Directional Spreading | Mean Directional Spreading | ° |
|---|---|---|
| Surface Ice Concentration | Surface Ice Concentration, given as ratio of ice on total surface (from ERA5) | - |
| 1st peak significant wave height | Significant wave height of the largest spectral peak | m |
| 1st peak mean period $T_{0-1}$ | Mean period of the largest spectral peak | s |
| 1st peak mean direction | Mean direction of the largest spectral peak | °N |
| 1st peak directional spread | Directional spread of the largest spectral peak | ° |
| 2nd peak significant wave height | Significant wave height of the 2nd largest spectral peak | m |
| 2nd peak mean period $T_{0-1}$ | Mean period of the 2nd largest spectral peak | s |
| 2nd peak mean direction | Mean direction of the 2nd largest spectral peak | °N |
| 2nd peak directional spread | Directional spread of the 2nd largest spectral peak | ° |
| 3rd peak significant wave height | Significant wave height of the 3rd largest spectral peak | m |
| 3rd peak mean period $T_{0-1}$ | Mean period of the 3rd largest spectral peak | s |
| 3rd peak mean direction | Mean direction of the 3rd largest spectral peak | °N |
| 3rd peak directional spread | Directional spread of the 3rd largest spectral peak | ° |

## Model performance evaluation: sea surface height

Modelled Sea Surface Height (SSH) was compared globally versus along-track altimetry data from the Copernicus Marine Service (CMS) (Pujol, 2022), which comprises several altimeter missions (Topex-Poseidon, Jason-1, OSTM/Jason-2, Sentinel-3A, ERS-1, ERS-2, Envisat, Geosat Follow On, Cryosat, SARAL/AltiKa, HY-2A) from 1993 ongoing. The dataset includes all the atmospheric forcing contributions to sea level variation, included as a linear component to the overall Sea Level Anomaly (SLA). The model output was space-time interpolated to the corresponding points along the tracks. The paired observation and modelled values were then binned on a validation grid with a spatial resolution of 0.5°x0.5° and the skill was evaluated for each cell.

In addition to the SSH validation at the open ocean, the performance of the nearshore SSH was assessed versus the tidal gauge data provided by the GESLA-3 database (Woodworth et al., 2016; Haigh et al., 2022).

As our barotropic model is unable to reproduce low-frequency variability of sea levels and Sea Level Rise, both simulations and observations of SSH were detrended prior to comparison.

The statistical indicators used to quantify the skill both of the offshore and nearshore SSH are:

- Root Mean Squared Error (RMSE):

$$RMSE = \sqrt{\frac{\sum_{i=1}^{N}\left(\eta_i^m - \eta_i^o\right)^2}{N}} \quad , \tag{1}$$

where N is the number of observation-simulation pairs, and $\eta_i^m$ and $\eta_i^o$ are the modelled and observed SSH, respectively.

- Percentage Root Mean Squared Error, RMSE(%), similar to RMSE but normalized by the difference between the 1st and 99th percentile of the observed water levels:

$$RMSE(\%) = \frac{\sqrt{\frac{\sum_{i=1}^{N}\left(\eta_i^m - \eta_i^o\right)^2}{N}}}{p_{99}(\eta^o) - p_1(\eta^o)} \quad , \tag{2}$$

- Pearson correlation:

$$r = \frac{\sum_{i=1}^{N}(\eta_i^m - \overline{\eta_m})(\eta_i^o - \overline{\eta_o})}{\sigma_m \sigma_o} \quad , \tag{3}$$

where $\overline{\eta_m}$ and $\overline{\eta_o}$ are the mean modelled and observed SSH, and $\sigma_m$ and $\sigma_o$ are the standard deviations of the 2 signals.

To assess the behavior of the model in normal and extreme water level conditions, the skill indicators were evaluated a) on the whole time series and b) on the data beyond the 95[th] percentile of the model or observation of each cell/tidal gauge.

## Model performance evaluation: waves

The modelled significant wave height $H_s$ was validated with the along-track altimetry $H_s$ data provided by CMS, in collaboration with the Climate Change Initiative (CCI) of the European Space Agency (ESA) (Dodet et al., 2020). Similar to SSH, the model data were space-time interpolated to the points along the satellite tracks, and binned on the same validation grid. $H_s$ was also compared with buoy data from a dataset of wave buoys from different sources, including the National Buoy Data Center (NDBC) and the European Marine Observation and Data Network (EMODnet).

The statistical indicators used to quantify the skill of significant wave height are:

- Normalized Bias (NBI):

$$NBI = \frac{\sum_{i=1}^{N} H_{s_i}^m - H_{s_i}^o}{\sum_{i=1}^{N} H_{s_i}^o} \quad , \tag{4}$$

where $H_{s_i}^m$ and $H_{s_i}^o$ are the modelled and observed significant wave height, respectively.

- The Root Mean Squared Error (RMSE) with a definition similar to that of SSH:

$$RMSE = \sqrt{\frac{\sum_{i=1}^{N}\left(H_{s_i}^m - H_{s_i}^o\right)^2}{N}} \quad . \tag{5}$$

- Normalized Root Mean Squared Error (NRMSE):

$$NRMSE = \sqrt{\frac{\sum_{i=1}^{N}\left(H_{s_i}^m - H_{s_i}^o\right)^2}{\sum_{i=1}^{N}\left(H_{s_i}^o\right)^2}} \quad . \tag{6}$$

Similar to SSH, the behavior of the model was assessed both in normal and extreme conditions of $H_s$.

# Results

The model shows overall good skill of SSH versus altimeters, with an average RMSE of about 7.9 cm, a RMSE(%) of 17.47%, and a Pearson correlation of 0.55. In large portions of the global ocean RMSE(%) is below 10% and the correlation is > 0.7 (Figure 1). The model performance decreases in areas characterized by low-frequency variability, for example in correspondence of the western-boundary currents. This is clear at the Gulf Stream, the Kuroshio, the Brazil current, at the Mozambique current and the gradient between the latter and the Southern Atlantic and Southern Indian currents, and the Eastern Australia current, especially when the model skill is evaluated in terms of correlation (Figure 1b). Model performance also varies with latitude. At higher latitudes both north- and southward correlation is higher and RMSE(%) lower, whereas closer to the equator the correlation becomes non-significant, despite RMSE(%) values in the order of 20% or lower (Figure 1). For several semi-enclosed basins, such as the Mediterranean Sea, the Red Sea, the Persian Gulf, the East China Sea, part of the South China Sea, the Gulf of Thailand, the Arafura Sea, the correlation is significantly higher than in open-ocean areas at the same latitude. Higher values beyond the 95th percentile of SSH are to be expected as the normalization term is the same as for the whole signal. The map of RMSE(%) beyond the 95th percentile presents spatial patterns similar to the ones found for the whole signal, but with higher relative uncertainty at the western boundary currents and at some section of the Southern Circumpolar Current (Figure 2a). The correlation model-altimeters beyond the 95th percentile shows a general improvement over the whole signal, with r > 0.9 in large swaths of the domain, and significant values of the correlation even in proximity of the equator (Figure 2a).

The comparison of simulated SSH with tidal gauge records shows similar model performance compared to altimeters. There is overall good skill, especially at higher latitudes for both mean (Figure 3) and extreme conditions (Figure 4), and also for tidal gauges located in proximity to the coast.

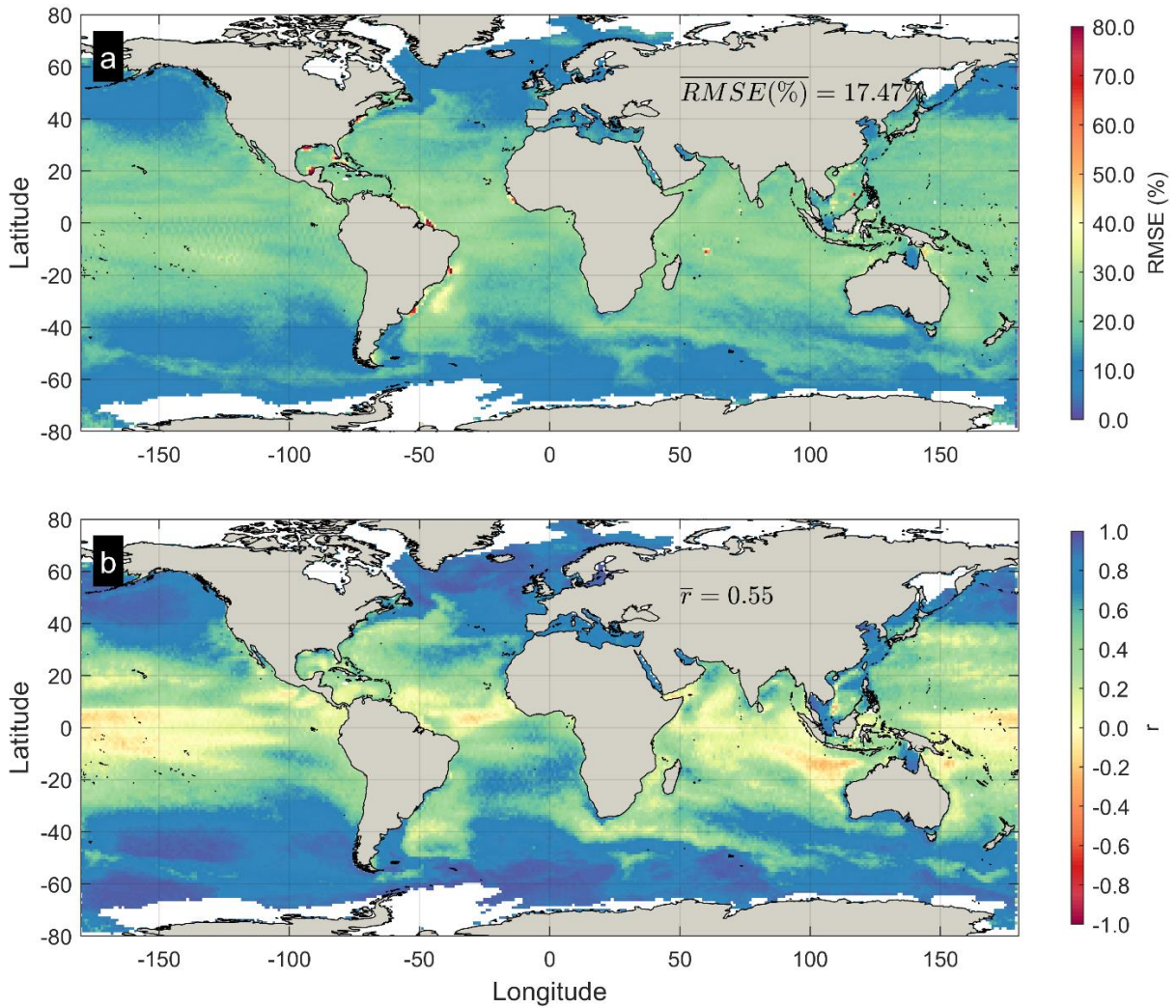

Figure 1: Skill for Sea Surface Height (SSH), model vs altimeters, a: percentage Root Mean Square Error RMSE(%), b: Pearson Correlation.

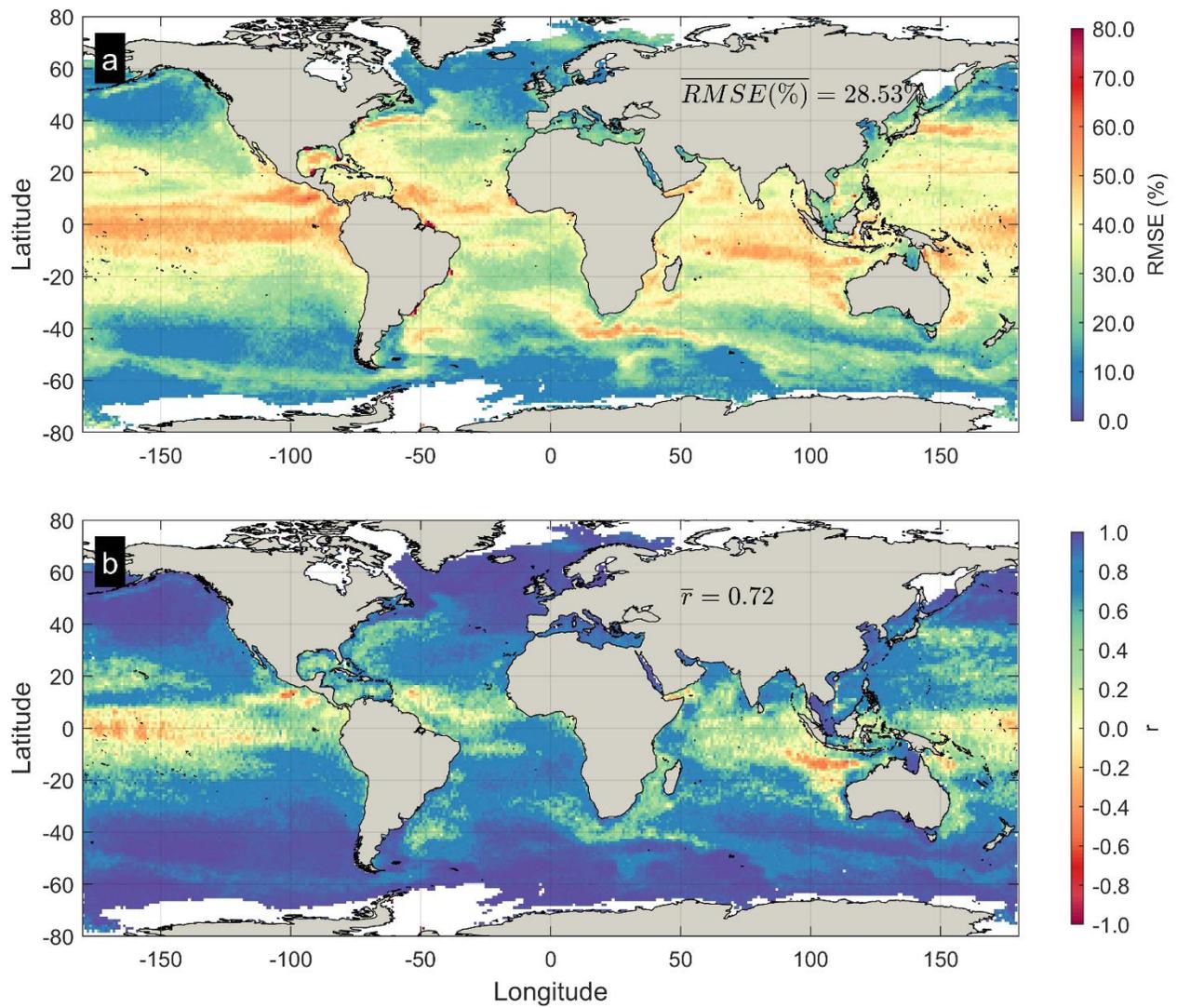

Figure 2: As in Figure 1, but for SSH beyond the 95th percentile of altimeters or model.

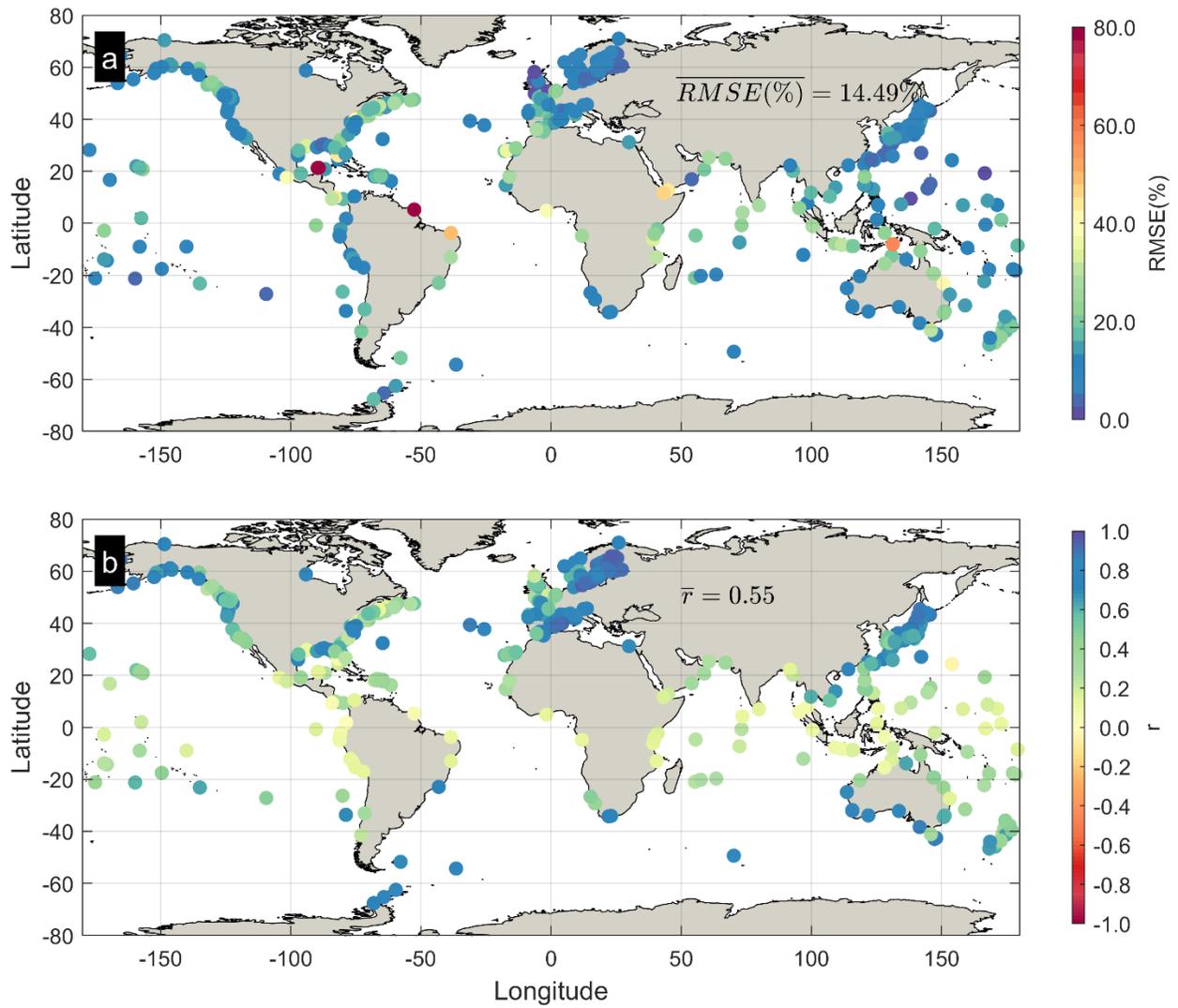

*Figure 3: Skill for Sea Surface Height (SSH), model vs tidal gauges, a: percentage Root Mean Square Error RMSE(%), b: Pearson Correlation*

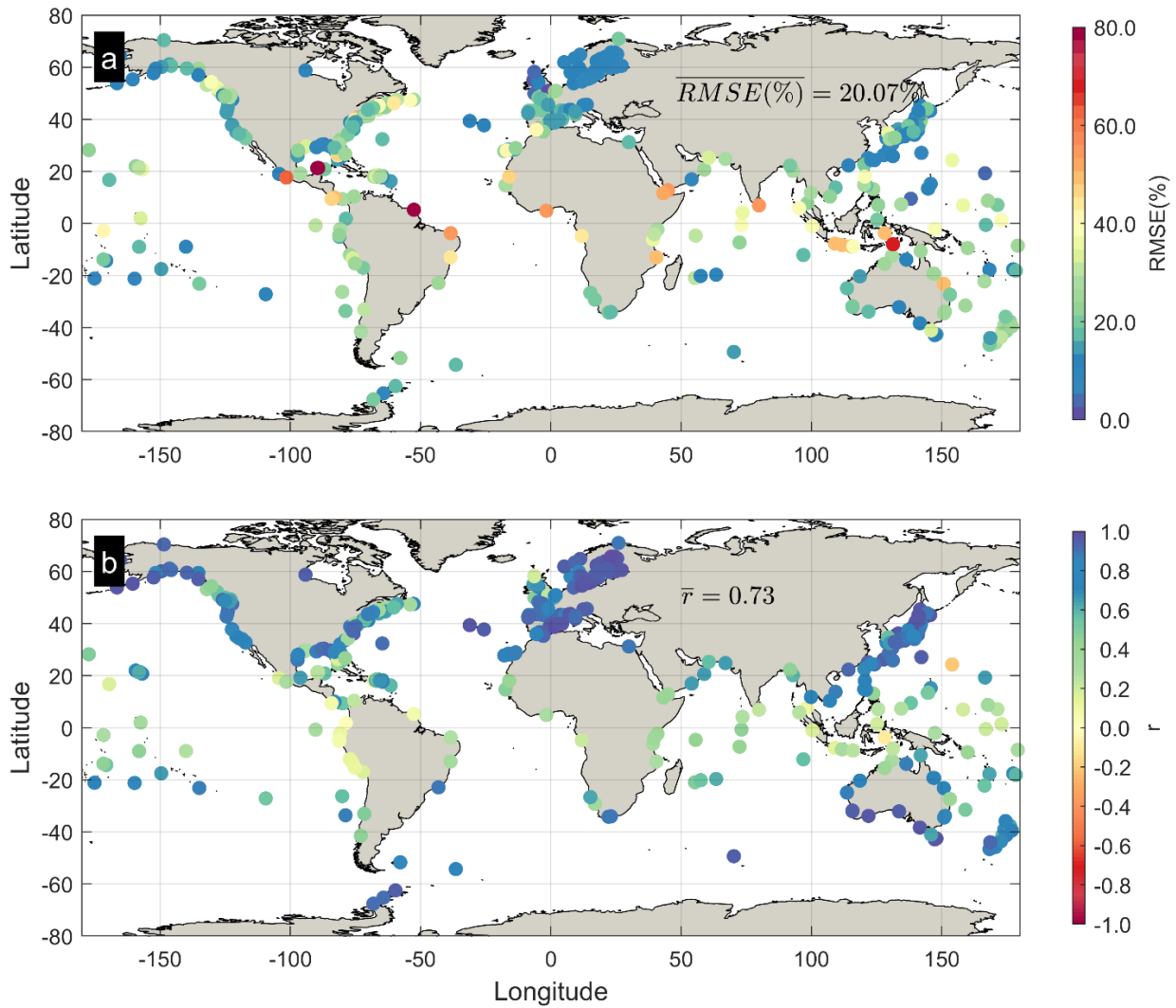

*Figure 4: As in Figure 3, but for SSH beyond the 95th percentile of gauges or model.*

The comparison of modelled vs altimeter values for $H_s$ shows that NRMSE is smaller than 25% over most of the domain (Figure 5a). Contrary to the skill for SSH (Figure 1a), NRMSE increases for higher latitudes north and south and is smallest closer to the equator. Biases are generally positive at low latitudes and negative at high latitudes, especially in the Northern Hemisphere but also in large swaths of the Southern Hemisphere (Figure 5b). The mean normalized bias over the entire domain is -3.73%. When the whole wave climate is considered (both mean and extreme conditions), the bias is generally negative in semi enclosed basins, such as the Mediterranean Sea, the Red Sea, the Persian Gulf, the East China Sea, part of the South China Sea, the inner seas of the Indonesian archipelago. Part of the Antarctic coasts are instead characterized by a positive bias (Figure 5b). The evaluation of the model skill on the extreme tails ($H_s$ beyond the 95th percentile of the observations) shows patterns of NRMSE similar to those found for the whole signal, but with significant improvements in some areas, in particular in semi-enclosed basins (Figure 6a). Beyond the 95th percentile the bias of Hs becomes more negative than in mean conditions, -7.85% on average, and the only areas retaining a positive bias are Coral Sea, North-West of Australia, and parts of the western Indian Ocean. The bias is low especially at high latitudes in the Northern Hemisphere, with negative bias values less than -15% in the Labrador Sea, in the Hudson Bay, along the coasts of Greenland and in other areas partially covered in ice during winter. In small, semi-enclosed basins such as the Mediterranean Sea, the Red Sea, the Persian Gulf, the inner seas of Indonesia and others, the negative bias improves beyond the 95th percentile of Hs compared with mean conditions (Figure 6b). The comparison of $H_s$ with buoys offers a picture similar to altimeters, though with more local variability as the ability of the model to reproduce observations can depend on the resolution of coastal micro-features and sheltering (Figure 7 and Figure 8). With few exceptions, the model generally shows skill in line with that versus altimeters even for buoys close to the coasts, though with a significantly more pronounced negative bias (-11.8% on the whole signal, and -16.26 beyond the 95th percentile, Figure 7b and Figure 8b).

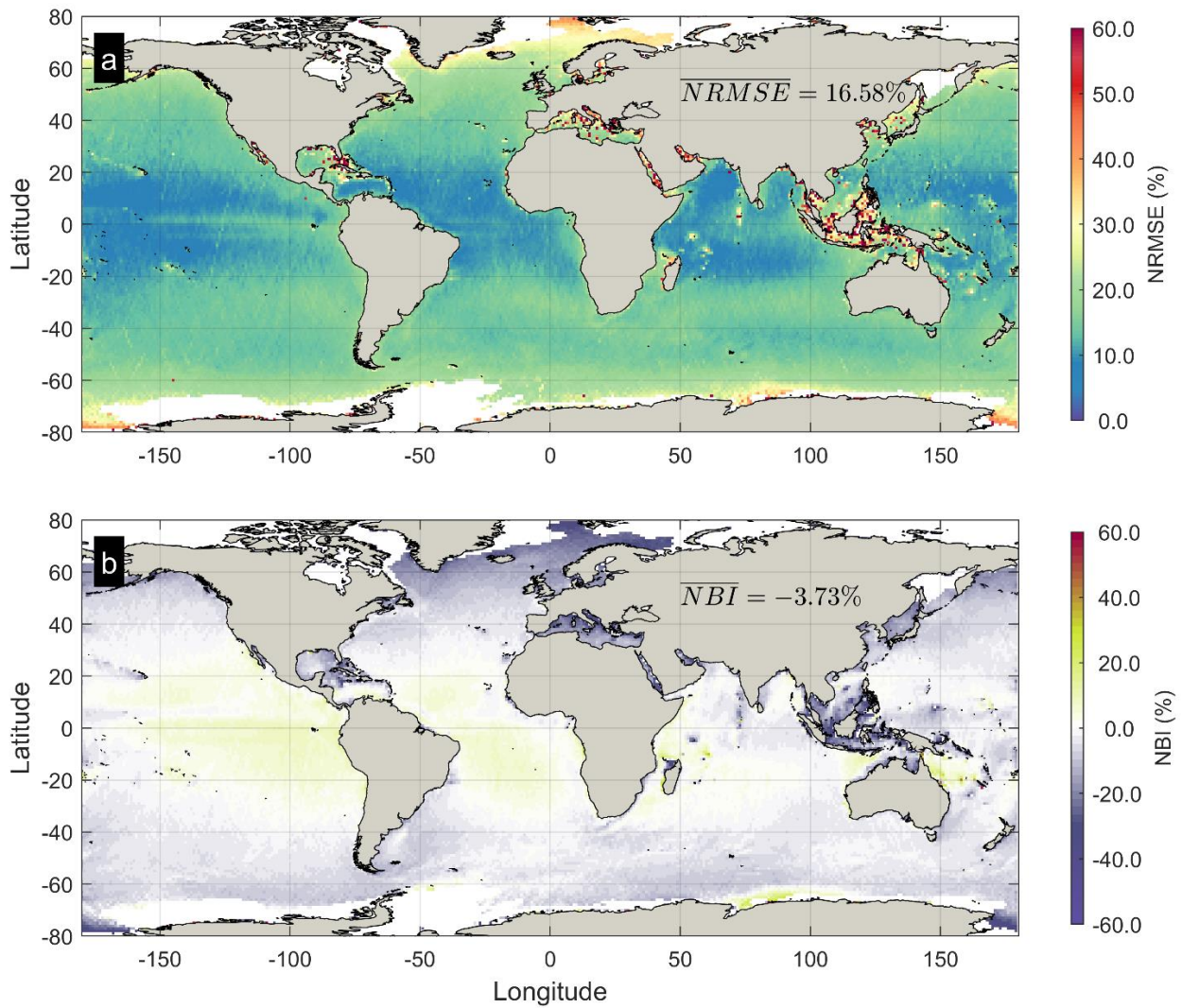

Figure 5: Skill of Significant Wave Height ($H_s$), model vs altimeters, a: percentage Normalized Root Mean Squared Error (NRMSE) b: Normalized Bias (NBI).

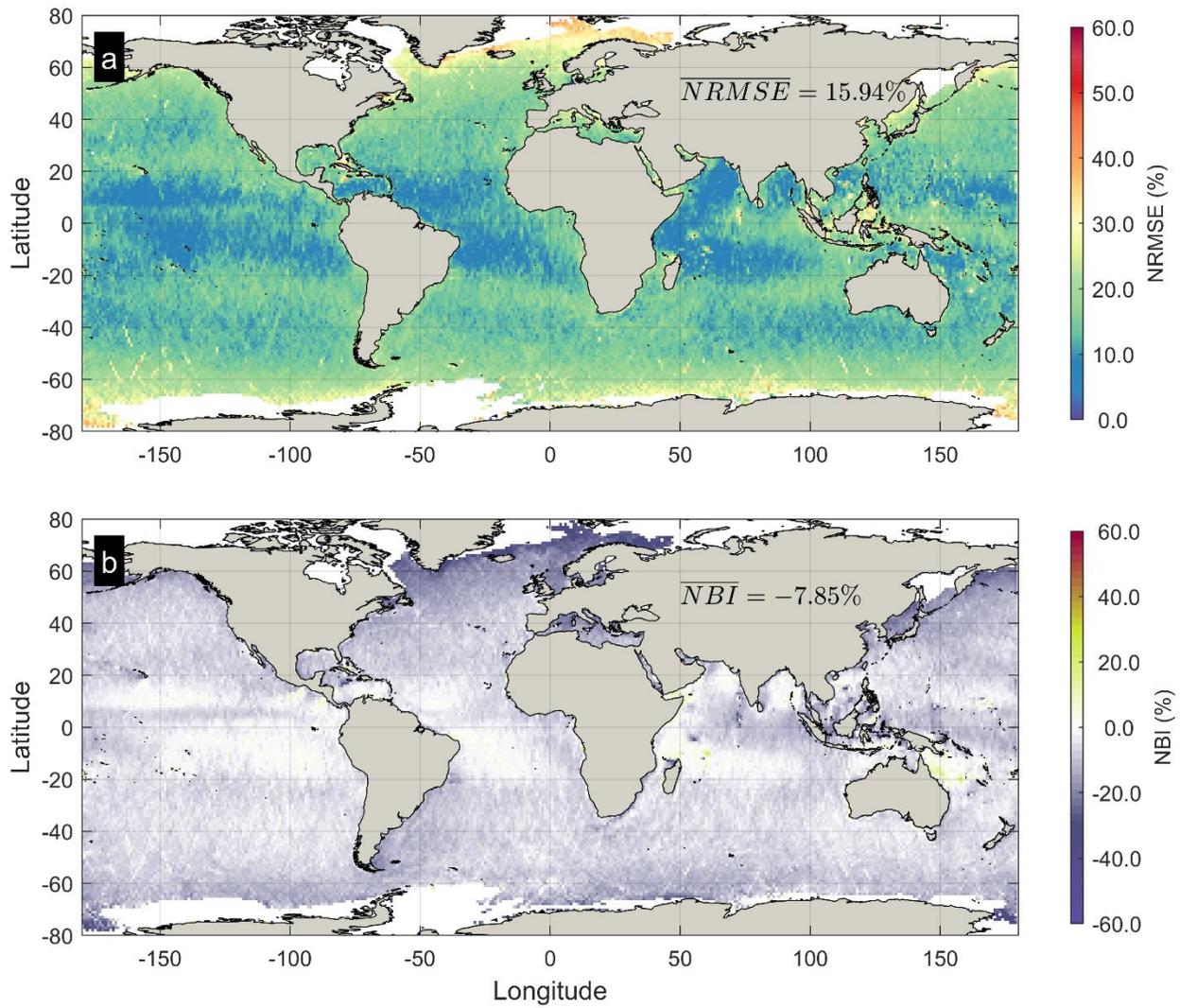

*Figure 6: As in Figure 5, but for H_s beyond the 95th percentile of altimeters or model.*

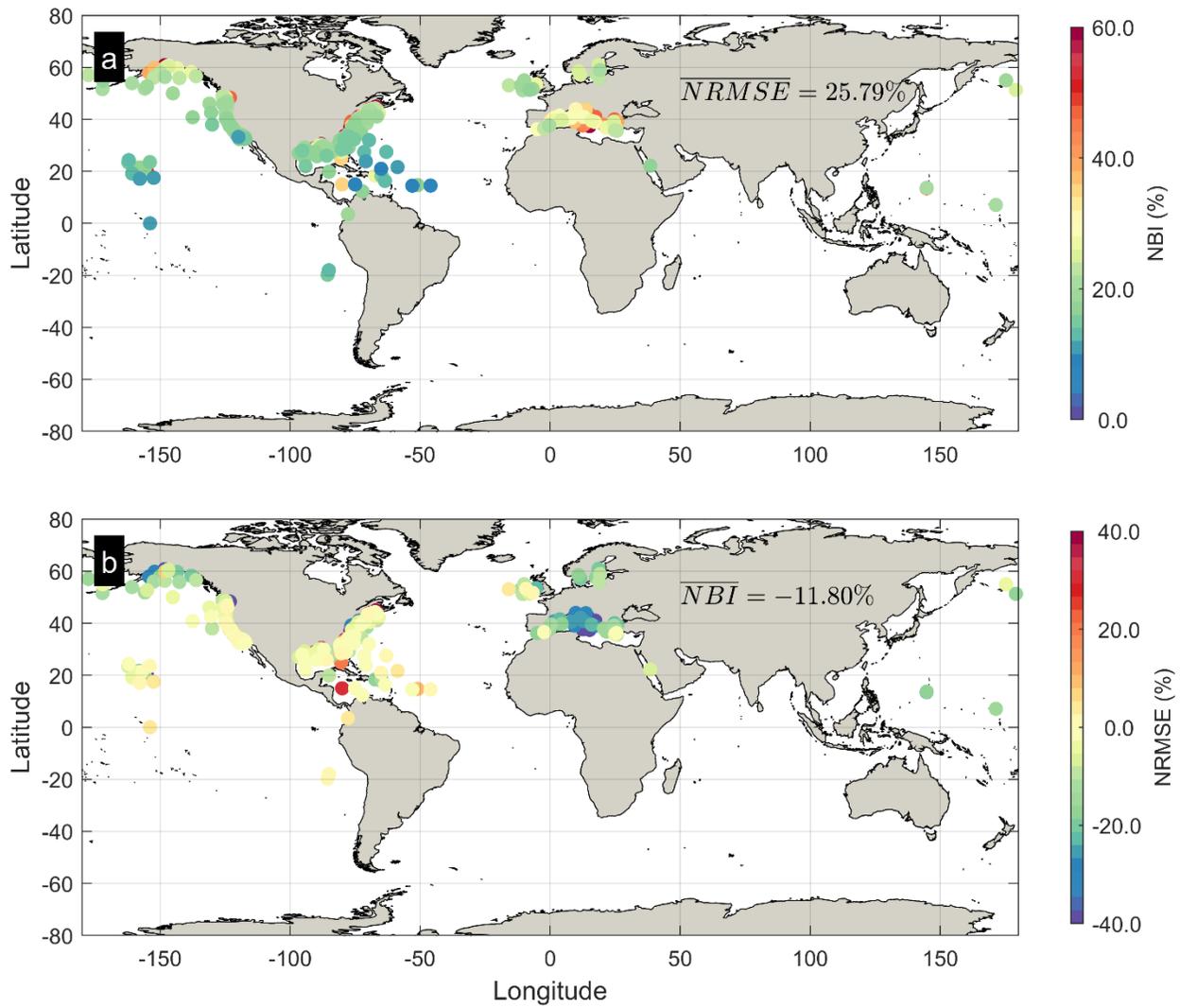

*Figure 7: Skill of Significant Wave Height (H$_s$), model vs buoys, a: percentage Normalized Root Mean Squared Error (NRMSE) b: Normalized Bias (NBI).*

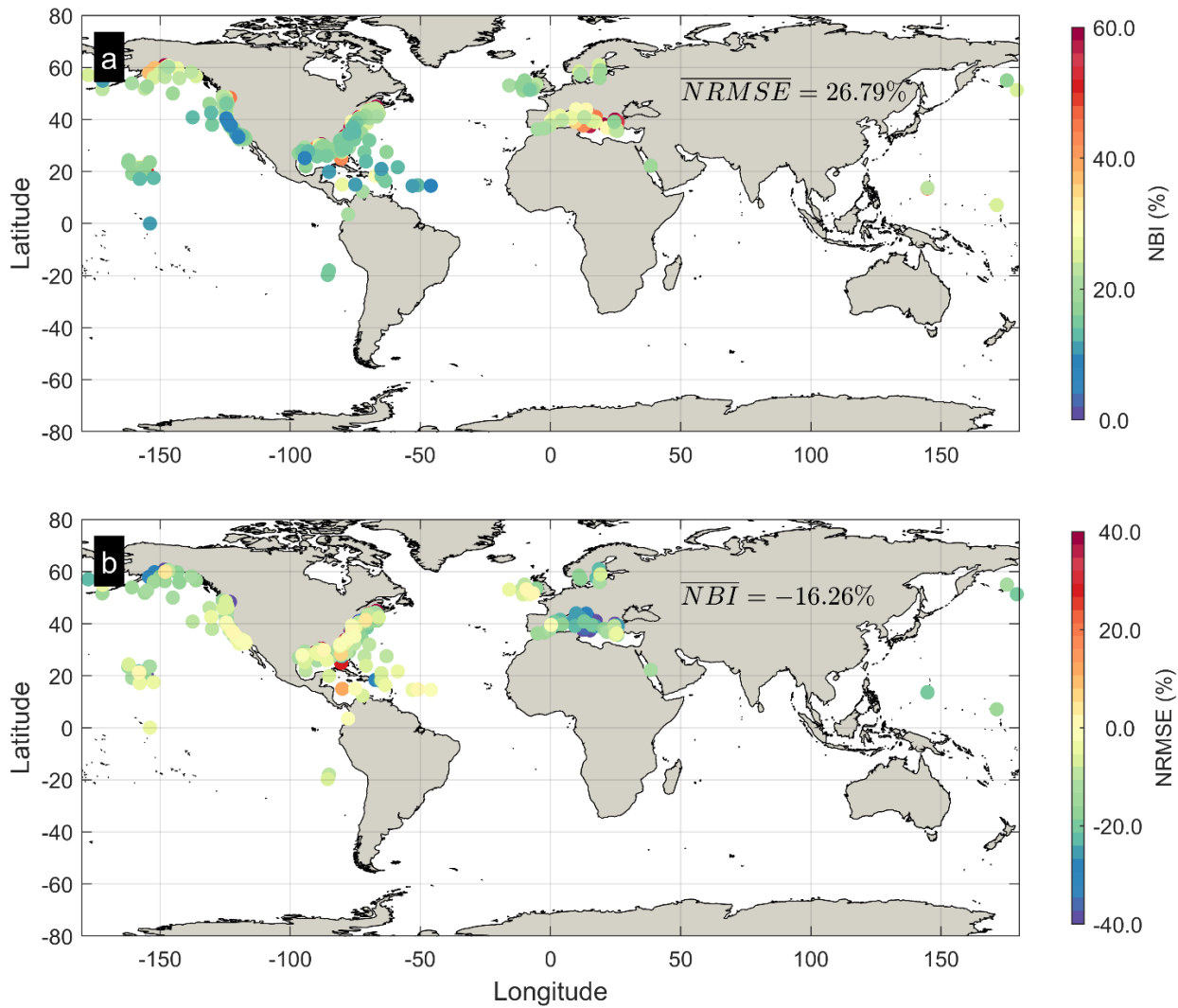

*Figure 8: As in Figure 7, but for H$_s$ beyond the 95th percentile of buoys or model.*

## Discussion

### Offshore skill

Overall, our results show offshore skills in line with the ones of other contemporary large-scale models, for both storm surges (Muis et al., 2016; Vousdoukas et al., 2017; Fernández-Montblanc et al., 2020; Tadesse et al., 2020) and waves (e.g. Mentaschi et al., 2020; Morim et al., 2022; Perez et al., 2017; Smith et al., 2021).

The offshore skill of SSH shows a clear dependence on latitude (Figure 1, Figure 2), with values of RMSE(%) and correlation significantly better at mid-high latitudes with respect to tropical and equatorial areas. This is a common feature in storm surge models and is due to a range of factors. The variability of sea levels at mid-high latitudes is dominated by the eddy activity and baroclinic instability in the atmosphere, which result in short-term inverse barometric effect and wind setup that our model can capture. The lower atmospheric eddy activity in tropical/equatorial regions is a reason for the worse skills of SSH in these areas (Hoskins and Valdes, 1990). At low latitudes storm surges are typically caused by tropical cyclones, hurricanes, and typhoons, which are relatively rare events, with a limited contribution to the mean signal. Though ERA5 can reproduce such events to some extent, limitations still exist, partially related to the still relatively low spatial resolution (Hersbach et al., 2020), with results usually characterized by negative biases of wind. Another reason is the complex ocean currents and coastal geography in the equatorial region, which is characterized by weak Coriolis acceleration. The equatorial region is home to the Equatorial Undercurrent (EUC) and the Equatorial Counter Current (ECC), which are difficult to model due to their strong variability and complexity (e.g. Stellema et al., 2022). These currents can also cause upwelling and vertical circulation, which can have an impact on SSH, but the ability of barotropic models to reproduce it is limited. Related to this is the low-frequency variability that characterize the sea levels in tropical and equatorial areas, due to the Tropical Instability Waves and the seasonal cycle, and the associated thermostatic expansion. Another source of uncertainty is the oscillating intensities of the monsoons and of the Walker cells, and the associated patterns of variability such as the El Nino Southern Oscillation (Nidheesh et al., 2013). Other areas characterized by low-frequency variability are the western boundary currents, where the water levels depend on currents' intensity. The better skill in relatively sheltered tropical/equatorial basins such as the Red Sea, the Persian Gulf, the East China Sea, part of the South China Sea, the Gulf of Thailand, the Arafura Sea, is likely due to the relatively smaller low-frequency variability.

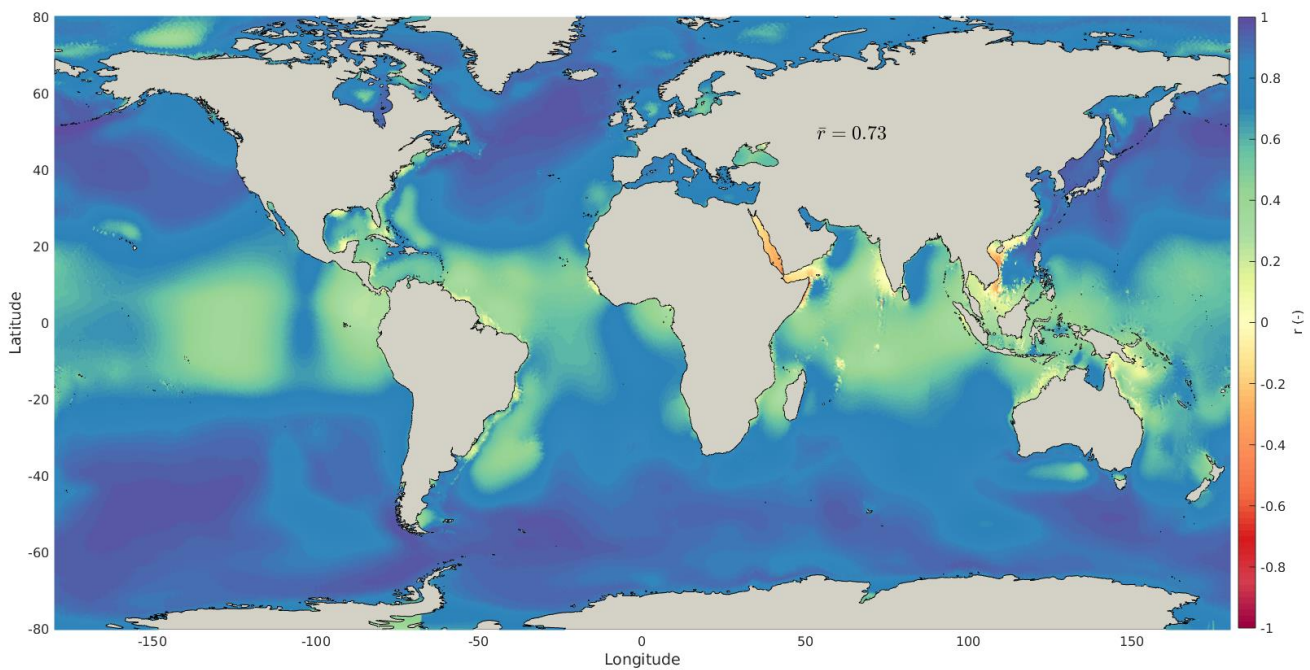

*Figure 9: Comparison of model SSH versus Dynamic Atmospheric Correction (DAC) from (Pujol, 2022). Map of correlation coefficient r.*

That low-frequency variability is a major contribution to the error is confirmed by a comparison of our model's SSH against the data of Dynamic Atmospheric Correction (DAC) from Pujol, (2022), which better identifies the short-term variability than SLA (Figure 9). The comparison, done for the mean signal, shows a significantly improved model skill, with a mean correlation of 0.73 versus 0.55 against SLA (Figure 1b), and better performances everywhere, with the exception of the Red Sea and the coasts of Vietnam.

Wave results show overall good skill, though with a tendentially negative bias, especially low in areas characterized by seasonal presence of ice, e.g. in the southern portion of the Arctic Ocean, in the Labrador Sea, and in the Southern hemisphere in the Weddell and Ross seas, and in semi-enclosed basins, with consequent higher values of NRMSE (Figure 5). The error in $H_s$ in these areas is related to the high uncertainty in the sea ice concentration datum provided by ERA5, especially in the Marginal Ice Zones (Renfrew et al., 2021), together with the fact that the presence of ice invalidates the altimetric observations of $H_s$ (Dodet et al., 2020). On the other hand, introducing a parameterization of ice-induced wave dampening comes with benefits, as simulations neglecting it result in significant positive biases of $H_s$ in the Southern Ocean and in the Northern Pacific Ocean and Atlantic Ocean that are now absent. The negative bias and NRMSE increase beyond the 95th percentile (Figure 6), reflecting a general tendency of our wave models in underestimating the peaks, partly due to an underestimation of extreme winds in ERA5 (e.g. Campos et al., 2022) as well as possible limitations of 3rd generation models in extreme conditions (Cavaleri, 2009). At the same time, in enclosed and semi enclosed basins (the Mediterranean Sea, the Red Sea, the Persian Gulf, the inner seas of Indonesia and others) the negative bias of $H_s$ beyond the 95th percentile improves with respect to the whole signal. This suggests that the higher resolution of ERA5 compared to previous reanalyses and that of our model allow for a better reproduction of mesoscale events relevant for the statistics of the extremes in short fetches (e.g. Mentaschi et al., 2015).

### Nearshore skill

The advantage of the high resolution used in this study becomes clear when examining the model skill vs tidal gauges and buoys located close to the coasts. Here our model generally outperforms lower resolution ones, in particular the ones from previous studies by the authors (Mentaschi et al., 2017; Vousdoukas et al., 2017, 2018). At global scale, the performance of SSH versus tidal gauges is remarkably in line with the one versus altimeters. The global mean RMSE(%) / correlation is 17.09 % / 0.58 versus altimeters, and 14.49 % / 0.54 versus tidal gauges (Figure 1, Figure 3). This picture is consistent for SSH beyond the 95th percentile, with a global mean RMSE(%) / correlation of 27.41 % / 0.74 versus altimeters, and 20.07 % / 0.73 versus tidal gauges (Figure 2, Figure 4).

For significant wave height the skill versus buoys is significantly worse than that versus altimeters, with a more pronounced negative bias, especially beyond the 95th percentile. When considering the whole signal, the global mean NRMSE / NBI is 16.58 % / -3.73 % versus altimeters, and 25.79 % / -11.8 % versus buoys (Figure 5, Figure 7). Beyond the 95th percentile the figures are 15.94 % / -7.85 % versus altimeters, and 26.67 % / -16.26 % versus buoys (Figure 6, Figure 8). The NRMSE versus buoys can be decomposed into its random and systematic components according to (Mentaschi et al., 2013), revealing that beyond the 95th percentile the bias account for roughly 20% of the mean squared deviation. The larger bias of $H_s$ nearshore is, beside the underresolved coastal zone, likely due to multiple factors, such as limitations in reproducing growth-dissipation in short fetches, and uncertainties in the parameterizations of shallow-water dynamics.

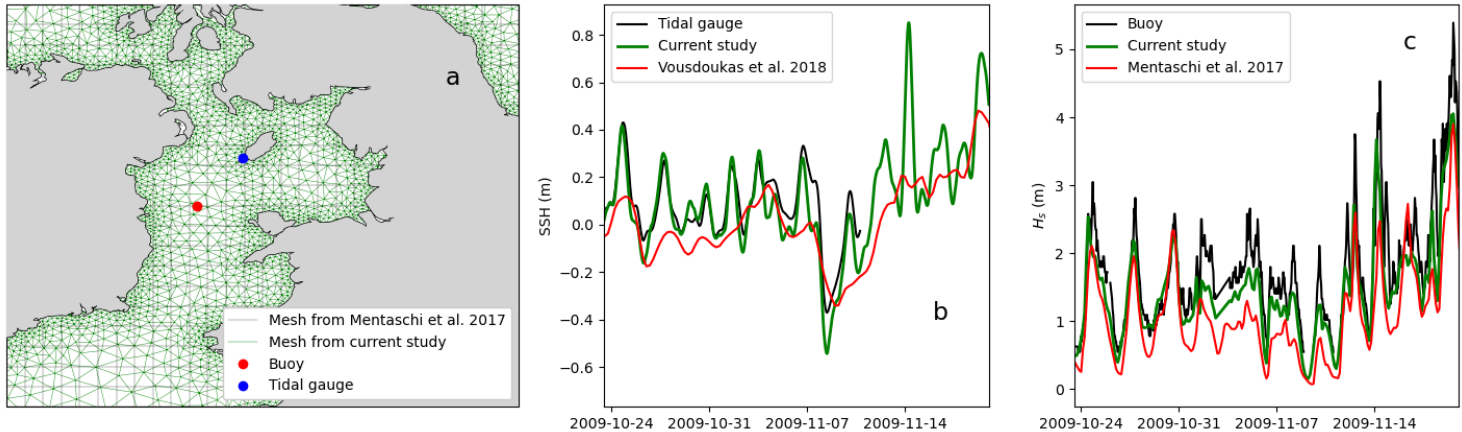

*Figure 10: Time series of SSH at a tidal gauge (b) and of $H_s$ at a buoy (c) in the Irish Sea (a). In map (a) the mesh used in the current study (green line) is superimposed to the 0.25° subdomain used for Northern Europe in (Mentaschi et al., 2017), and the positions of the buoy and the tidal gauge are represented. In the time series panels (b and c) the black line represents the observations, the green line the results from the present study, the red line for Hs the results from Mentaschi et al., (2017) and for SSH from Vousdoukas et al., (2018).*

The significant improvement over lower resolution, non-coupled models for both SSH and $H_s$ relates to the ability to better reproduce the dynamic effect of small coastal features, and to considering the interactions between waves and currents, and the effect of waves on the sea surface roughness. A comparison of the new hindcast versus previous ones (Mentaschi et al., 2017; Vousdoukas et al., 2018, Figure 10) shows how the present model better captures local short-term variability. For the 20% of tidal gauges with the best RMSE for SSH in the present study, the old hindcast showed nowhere a better skill (Figure 11a). Furthermore, the RMSE/RMSE(%) of SSH was larger than the 80[th] percentile of RMSE/RMSE(%) of the present study in >50% of the tidal gauges. If SSH beyond the 95[th] percentile is considered, this percentage exceeds 60% (Figure 11ab). For $H_s$, in normal conditions the improvement of skill is in line with that of SSH, while if we consider $H_s$ beyond the 95[th] percentile the improvement of skill with respect to the old hindcast is still clear, but less marked, due to the significant negative bias of $H_s$ (Figure 11cd). In this respect, it is worth mentioning that the setup of the physics in WW3 and D-FLOW in (Mentaschi et al., 2017; Vousdoukas et al., 2018) is similar to that of the present study, and the offshore skill comparable. Therefore, the nearshore improvement of skill is due to the increase in resolution and to the coupling.

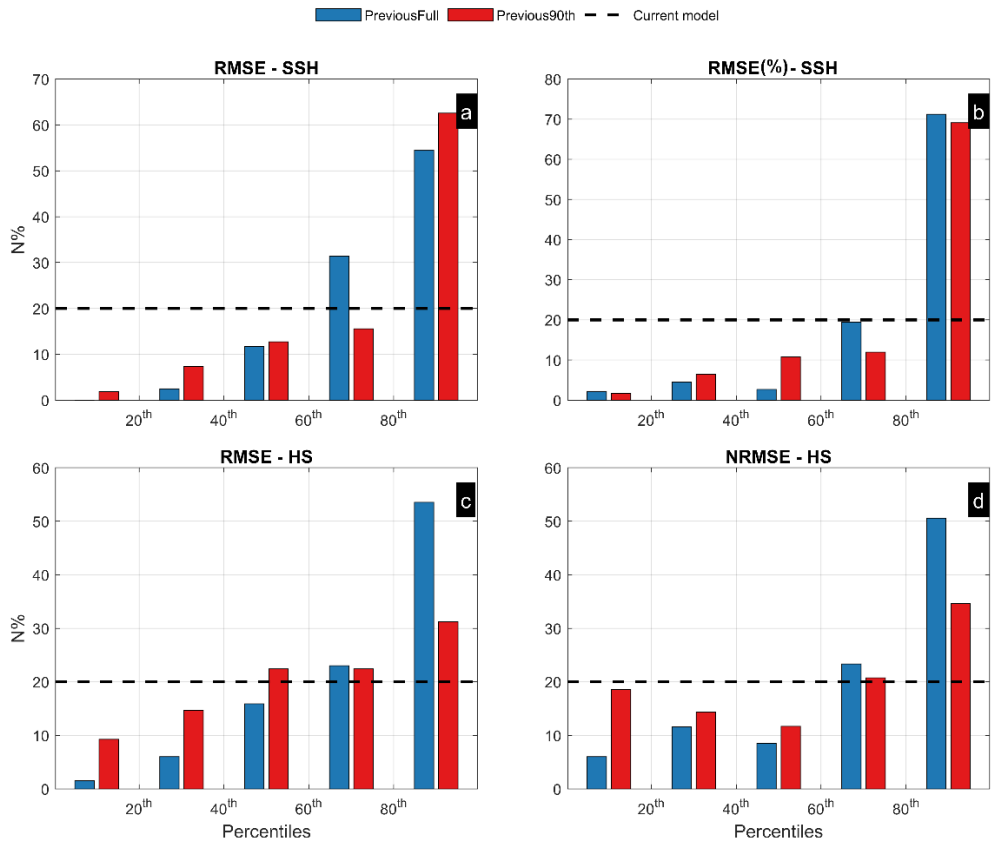

*Figure 11: Skill comparison between the present study and previous, lower-resolution hindcasts* (Mentaschi et al., 2017; Vousdoukas et al., 2018). *In the panels, the tidal gauges (ab) and the buoys (cd) are grouped by quintile of error in the present study. The dashed line in each panel reproduces the constant percentage of gauges/buoy falling in each quintile. The bars represent the percentage of gauges/buoys of the older hindcasts falling in the same quantiles, in normal conditions (blue bars), and for the part of the signal beyond the $95^{th}$ percentile. The skill indicators are RMSE and RMSE(%) of SSH (ab), and RMSE and NRMSE of $H_s$ (cd).*

## Limitations

Our model represents a clear improvement on previous lower resolution hindcasts. However, it comes with a number of limitations that will be listed in this section. Tidal forcing was not included in the setup, as we aimed to isolate the contribution of the storm surges. While tide elevation can be added a-posteriori, from tidal databases such as FES2014 (Lyard et al., 2021) or TPXO (Dushaw et al., 1997), recent studies pointed to the importance on non-linear interactions between tides and storm surges nearshore (Krien et al., 2017; Fernández-Montblanc et al., 2019; Arns et al., 2020; Tausía et al., 2023). Such non-linear interactions are not explicitly accounted for in this study.

As a resolution of 2-4 km is not enough to resolve exhaustively everywhere coastal processes such as nearshore wave dissipation, we opted not to introduce the nearshore wave-circulation coupling described by (Bennis et al., 2011) and implemented in SCHISM by (Martins et al., 2022). Therefore, the water levels at the coast do not include the wave setup. This is a common approach for large-scale models, but it should be mentioned because it is important for certain applications.

The ability of our model to capture tropical cyclones (TCs) is still limited by the offshore grid resolution (~ 50 km) and by the ability of the forcing fields from ERA5 to reproduce TCs. This is another known limitation of large-scale models (e.g. Bloemendaal et al., 2019) and requires separate special modelling to be addressed.

As mentioned in the methods section, ERA5 forcing was selected for its advantages over other products (high spatial resolution for a global reanalysis, long time extent, ability to detect long-term trends). However, ERA5 comes with significant underestimation of extreme winds (e.g. Campos et al., 2022), which reflects in negative bias especially of extreme $H_s$.

Last but not least, this study was conducted using a 2D barotropic model for storm surges, which comes with a limited ability to reproduce 3D ocean circulation and currents. Though an accurate representation of 3D circulation is not an objective of this study, this could have an impact on storm surges, e.g., in relation to phenomena such as the Coriolis setup. Furthermore, this can have consequences on waves, related to the current-induced Doppler effect.

## Final remarks

In this study we developed the first global coupled hindcast of waves and storm surges on unstructured meshes, reaching a nearshore spatial resolution of 2-4 km, unprecedented for waves. Thanks to its high resolution, our hindcast achieves better skill nearshore than previous, lower resolution ones, in particular the ones developed in the past by the authors (Mentaschi et al., 2017; Vousdoukas et al., 2018). This offers several opportunities for advancements in understanding and quantification of coastal hazard and risk at large scale. For example, it offers the potential to better characterize areas of erosion or accretion, and reproduce the occurrence of past flooding events. Its extended time span (50 years, with its extension up to 70 years ongoing) offers opportunities for attribution studies in view of climate change.

The model and hindcast presented advance the state of the art on global scale coastal hydrodynamic modelling, and the limitations discussed represent some key challenges for future further improvement. Some of them can only be addressed in the long term, by running new models on different meshes using improved approaches. Among these, the obvious need for even higher spatial resolution to reproduce the coastal interaction between circulation and waves, the need of even better atmospheric forcing for more accurate modelling TCs and extremes in general, and the use of 3D circulation models to better reproduce ocean circulation.

Other limitations can be partially overcome via Machine-Learning post-processing of our results. Noticeably, the systematic errors of key quantities, such as SSH and $H_s$, can be reduced via bias correction versus long-term altimetric observations, similar to what is done with simulated climate variables by general circulation and regional climate models (e.g. Lange, 2019). The non-linear tide-surge interactions can be inferred from model runs that consider both, and our results of SSH can be corrected accordingly (Arns et al., 2020). Separate, high-resolution models can be run for TCs, for a better characterization of extreme conditions at lower latitudes, as done by (Vousdoukas et al., 2018). Wave runup can be parameterized using empirical approaches such as (Stockdon et al., 2006). These and other improvements will be the subject of follow-up research and will contribute to the creation of an even more accurate global database for coastal applications and engineering.

## Acknowledgements

The authors acknowledge Xavier Bertin for his useful suggestions and careful review of the manuscript.